\documentclass[notitlepage,twocolumn,amsmath,amssymb,aps]{revtex4-2}

\usepackage{physics}
\usepackage{amsmath}
\usepackage{xcolor}
\usepackage{graphicx} 

\newcommand{\oho}{ O-H...O }
 
\begin{document}

\renewcommand*{\bibfont}{\scriptsize}
 

\title{Deuteration removes quantum dipolar defects from KDP crystals}

\author{ Bingjia Yang }
\affiliation{ Department of Chemistry, Princeton University, Princeton, NJ 08544, USA}

\author{ Pinchen Xie }
\affiliation{ Program in Applied and Computational Mathematics, Princeton University, Princeton, NJ 08544, USA }

\author{ Roberto Car }
\affiliation{ Department of Chemistry, Department of Physics, Program in Applied and Computational Mathematics,
Princeton Materials Institute, Princeton University, Princeton, NJ 08544, USA }
 
\maketitle

\textbf{The structural, dielectric, and thermodynamic properties of the hydrogen-bonded ferroelectric crystal potassium dihydrogen phosphate ($\mathbf{KH_2PO_4}$), KDP for short, differ significantly from those of DKDP ($\mathbf{KD_2PO_4}$)~\cite{Samara1973,nelmes1987structural}. It is well established that deuteration affects the interplay of hydrogen-bond switches and heavy ion displacements that underlie the emergence of macroscopic polarization, but a detailed microscopic model is missing~\cite{tokunaga1987review, schmidt1987review}.  
Here we show that all-atom path integral molecular dynamics simulations can predict the isotope effects, revealing the microscopic mechanism that differentiates KDP and DKDP. Proton tunneling in the hydrogen bonds generates phosphate configurations that do not contribute to the polarization. These dipolar defects are always present in KDP, but disappear at low temperatures in DKDP, which behaves more classically. Quantum disorder confers residual entropy to ferroelectric KDP, explaining its lower spontaneous polarization and transition entropy relative to DKDP. Tunneling should also contribute to the anomalous heat capacity observed in KDP near absolute zero~\cite{lawless1982specific,foote1985low}. The prominent role of quantum fluctuations in KDP is related to the unusual strength of the hydrogen bonds in this system and should be equally important in the other crystals of the KDP family, which exhibit similar isotope effects~\cite{lines2001principles}. }

\begin{figure}[bht]
    \centering
    \includegraphics[width=\linewidth]{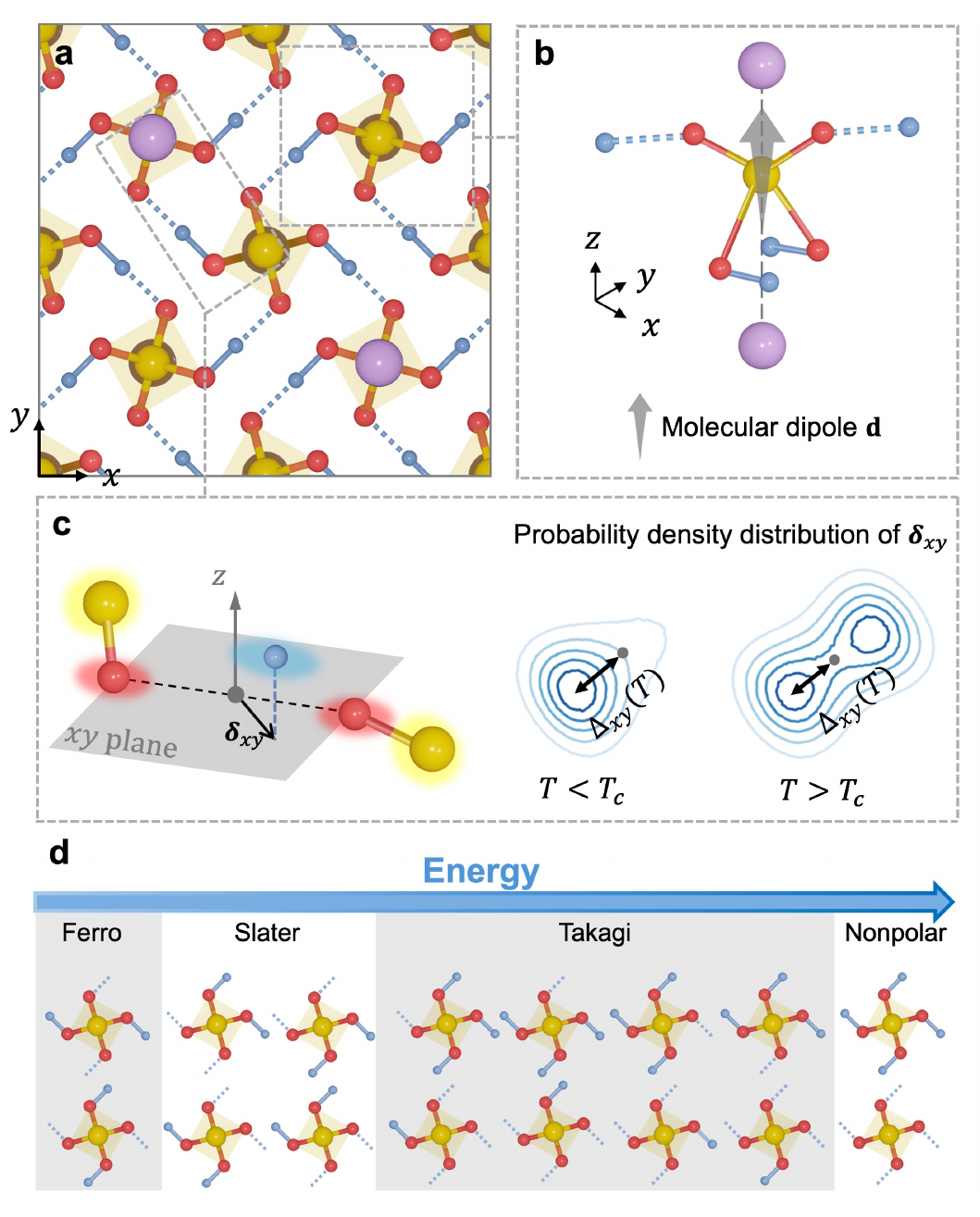}
    \caption{(a) Schematic representation~\cite{batoms} of the Fdd2 equilibrium structure of KDP. The z-axis is parallel to the $c$ crystallographic axis. 
    (b) Sketch of a ferroelectric KDP unit. K (shared), H (shared), O, and P ions are represented by purple, light blue, red, and yellow spheres. The grey arrow shows the direction of the molecular dipole moment $\mathbf{d}$.
    (c) Left: schematics of an O-H...O bridge. Colored shadows suggest the quantum delocalization of the particles. The geometric center of the oxygen pair is the origin of the coordinates. $\boldsymbol \delta_{xy}$ is the projection on the $xy$ plane of the position vector of the proton.  Right: contour plot of the probability density function $n(\boldsymbol \delta_{xy})$, below and above $T_c$, inferred from neutron diffraction experiments~\cite{nelmes1987structural, nelmes1987compilation}. Darker contour lines indicate higher probability density. $\Delta_{xy}(T)$, the distance of the maximum of $n(\boldsymbol \delta_{xy})$ from the origin, quantifies proton off-centering.
    (d) 16 types of KDP units are divided into four classes with decreasing energetic stability from left to right. Units in the same class have similar energetic stability. K ions and H-bonded H ions are not shown.}
    \label{fig:crystal}
\end{figure}

KDP is the prototypical member of a class of isomorphous hydrogen-bonded molecular crystals with formula unit $\mathrm{MH_2XO_4}$ (M=K, Rb, Cs, $\mathrm{NH_4}$; X=P, As)~\cite{lines2001principles}. The $\mathrm{XO_4}$ groups are held together by strong hydrogen bonds (H-bonds) (Fig.\ref{fig:crystal}(a)). The H-bonds are correlated with the polar distortion of the $\mathrm{XO_4}$ groups giving rise to ferroelectricity when M=K, Rb, Cs, and to anti-ferroelectricity when M=$\mathrm{NH_4}$. In these materials, $T_c$,
the paraelectric to ferroelectric (antiferroelectric) phase transition (PT) temperature, decreases with increasing pressure tending to zero at some critical pressure~\cite{lines2001principles}. When hydrogen is substituted with its heavier isotopes, changes of $T_c$ of the order of $100$K are observed~\cite{lines2001principles}. 
At ambient pressure, KDP and DKDP have $T_c$ equal to 123K and 229K, respectively. The isotope effects modulate piezoelectric, electro-optical, and nonlinear optical properties of KDP/DKDP~\cite{mason1946elastic, wang2017characteristics}, leading to different applications such as electro-optic modulators, frequency converters, and piezoelectric devices~\cite{baisden2016large, tressler1998piezoelectric}. Isotope effects on thermodynamic properties are a manifestation 
of nuclear quantum effects (NQE) in the atomic dynamics.      

Empirical observations indicate that NQE make strong H-bonds stronger~\cite{li2011quantum}, an effect that should favor more symmetric bond configurations in KDP than in DKDP. 
Structural changes across the PT can be related to two non-exclusive mechanisms:  displacive and order-disorder~\cite{rabe2007physics}.
In a displacive transition, spontaneous polarization emerges from an almost uniform structural distortion combining irreversible off-centering of the H/D ions with a polar distortion of the $\text{PO}_4$ groups induced by a displacement of the P ions as in Fig.\ref{fig:crystal}(b). In the
paraelectric phase, the H-bonds are uniformly symmetrized and the $\text{PO}_4$ groups are non-polar. 
The displacive mechanism underlies the tunneling model~\cite{blinc1960isotopic} and its variants~\cite{de1963collective, tokunaga1987review}, which are defined by effective Hamiltonians that treat the protons as two-level systems (TLS) and ignore or substantially reduce the degrees of freedom of the heavy ions (K, P, O). 
These models attribute the PT to the softening of a polar phonon mode combining collective proton tunneling with a polar distortion of the $\mathrm{PO_4}$ groups. The displacive mechanism was challenged by spectroscopic experiments~\cite{tokunaga1987review, TOMINAGA1983835,tokunaga1984light},
which found that, in the paraelectric phase, the \oho bridges are asymmetric but disordered, so that the $\mathrm{PO_4}$ groups are polarized in random directions with distortions not substantially weaker than in the ferroelectric phase. This scenario attributes spontaneous polarization to the establishment of long-range order, consistent with neutron diffraction experiments that show a lack of displacive-type structural distortions across the PT~\cite{nelmes1987compilation}. Indeed, the spatial distribution of the H/D ion relative to the neighboring O ions shows off-centered peaks in both ferroelectric and paraelectric phases, as illustrated in Fig.\ref{fig:crystal}(c). It is of great interest to understand precisely how NQE affect the disordering of the H-bond network. The insight gained on KDP should help understand the PT in the entire KDP family of crystals~\cite{lines2001principles}. It may also elucidate the isotope shifts larger than $100$K observed in hydrogen-bonded $\mathrm{H_2O_4C_4}$ and some $\mathrm{PbHPO_4}$-type crystals~\cite{mcmahon1990geometric, vrezina1975ferroelectricity}.

Key aspects of the order-disorder mechanism were elucidated by Slater~\cite{slater1941theory} and Takagi~\cite{takagi1948theory1, takagi1948theory2}, who pointed out that the local H-bond structure could be associated with the 16 phosphate configurations depicted in Fig.~\ref{fig:crystal}(d), which groups the phosphates into four classes, i.e., Ferro, Slater, Takagi, and Nonpolar units, in order of decreasing energy stability from left to right. The more stable Ferro and Slater units satisfy the ``ice rules'', because each phosphate has two donor and two acceptor H-bonds, in analogy with $\mathrm{H_2O}$ ice. The less stable Takagi and Nonpolar units break the ``ice rules'' as they have unequal numbers of donor and acceptor bonds. Attributing nominal charges to the ions and taking $\mathrm{K^{+}}$ into account, Ferro and Slater units ($\mathrm{H_2PO_4^{-}}$) satisfy local charge neutrality while Takagi and Nonpolar units violate it. 
Within electronic density functional theory (DFT), a molecular dipole moment $\mathbf{d}$ can be assigned to each phosphate group sharing H and K ions with its neighbors, as depicted in Fig.~\ref{fig:crystal}(b)~\cite{vanderbilt2018berry}. The local dipole $\mathbf{d}$ is defined in terms of the positions and charges of the ions (nuclei plus core electrons) 
and the centroids of the valence electrons (see Methods). The shared ions (K, H) carry half charge, the non-shared ions (P, O) and electronic centroids carry full charge. The product of polarization $\boldsymbol{ \mathcal{P}}$ and volume $V$ for a periodic supercell is equal to the sum of the dipoles $\mathbf{d}$ contained in that supercell modulo an immaterial constant that does not affect the changes of $\boldsymbol{ \mathcal{P}}$~\cite{resta2007theory, vanderbilt2018berry}. In view of its connection with the macroscopic polarization, the average $\mathbf{d}$ is a proper local order parameter for the PT. We find that Nonpolar units have nearly vanishing $\mathbf{d}$, Takagi and Slater units have non-zero $\mathbf{d}$ along directions approximately orthogonal to $z$, while Ferro units have non-zero $\mathbf{d}$ along $z$. For $T<T_c$, broken symmetry Ferro units dominate and the other units act as dipolar defects destabilizing the ordered phase. For $T>T_c$, extensive excitation of dipolar defects occurs. 
 \begin{figure*}[tbh]
    \centering
    \includegraphics[width=0.95\linewidth]{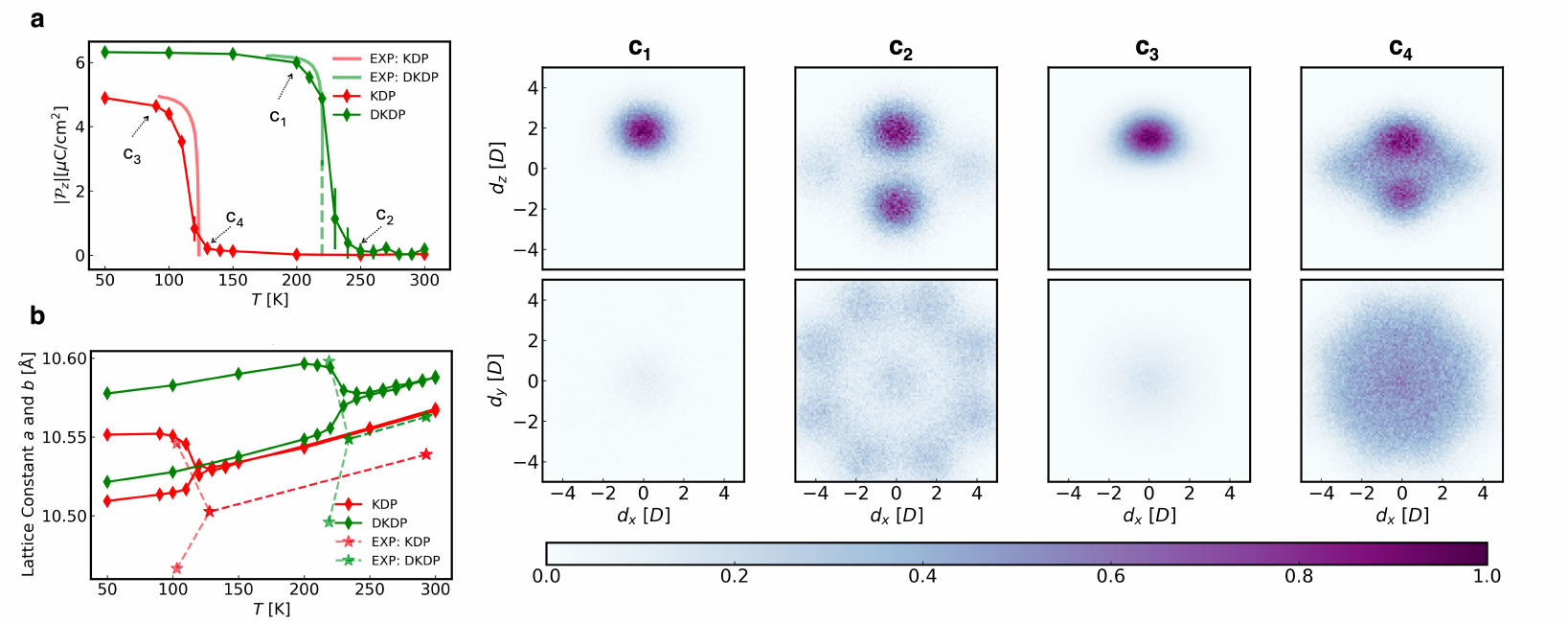}
    \caption{  (a) The longitudinal spontaneous polarization $|\mathcal{P}_z|$ as a function of $T$. The experimental data are for pure KDP and KD${}_{1.96}$H${}_{0.04}$PO${}_4$~\cite{Samara1973}. (b) Predicted and experimental~\cite{nelmes1987compilation} lattice constants $a$ and $b$ as functions of $T$. (c) The joint PDFs $n(d_x,0,d_z)$ (upper) and $n(d_x,d_y,0)$ (lower) for KDP and DKDP at the temperatures indicated in panel (a).
    }
    \label{fig:bonds}
\end{figure*}
Early models~\cite{schmidt1987review} explained semi-quantitatively the PT by reducing the degrees of freedom of each phosphate unit to the 16 discrete states of Fig.~\ref{fig:crystal}(d), ignoring the fluctuations that connect distorted configurations. At low temperatures, these fluctuations should be negligible in classical but not in quantum ferroelectrics.
Thermal and quantal fluctuations, including the effects of tunneling and anharmonicity~\cite{koval2002ferroelectricity, menchon2018ab}, can be simulated
with controllable accuracy with all-atom path-integral molecular dynamics (PIMD).
In this approach, $N$ distinguishable quantum ions are mapped onto $N$ classical ring polymers made of an equal number, $L$, of beads, subject to intra- and inter-polymer interactions. The intra-polymer interaction is harmonic and describes
the quantum delocalization of a particle, while the inter-polymer interaction, given by the potential energy surface of the system rescaled by $L$, describes the physical interactions of the ions. For large enough $L$, the equilibrium values of position-dependent observables estimated by sampling the ring polymer distribution reproduce the corresponding quantum values (see Methods).
Modeling the polarization evolution requires knowledge of the polarization surface, which gives the dependence of the polarization on the ion coordinates. This is possible with ab initio PIMD~\cite{marx1996ab}, an approach in which the potential energy and polarization surfaces are computed on the fly within Kohn-Sham DFT~\cite{kohn1965self}. So far only a small periodic system of four KDP units could be studied in this way~\cite{srinivasan2011isotope}, due
to the computational cost of ab initio PIMD. In these simulations, the accuracy
of the DFT approximation is crucial. Ref.~\cite{ menchon2018ab} found that semi-local functionals tend to overestimate the quantum delocalization of the hydrogens, destabilizing or even completely eliminating the ferroelectric phase, a difficulty likely associated to the self-interaction or delocalization error~\cite{perdew1981self, mori2008localization} of DFT approximations. Due to the exponential dependence of tunneling on the barrier height, a small underestimation of the latter can make the system a quantum paraelectric by removing the broken symmetry \oho configurations of the ferroelectric phase.

In this work, we overcome the limitations of ab initio PIMD and establish DFT-based models for KDP/DKDP that agree well with experiments. Specifically, we use neural network representations of the potential energy~\cite{zhang2018deep, wang2018deepmd} and polarization  surfaces~\cite{zhang2020deepwannier} (see Methods), to boost numerical efficiency and scalability.  The neural networks are trained on DFT data generated with the Strongly Constrained and Appropriately Normed (SCAN) functional approximation~\cite{sun2015strongly}.
Most importantly, we correct the excessive quantum delocalization of the H/D ions by stiffening the springs of the associated ring polymers, while leaving the other polymers unaffected. In practice, we multiply the spring constant $k = \frac{mL}{\beta^2\hbar^2}$, with $m$ the ion mass and $\beta$ the inverse temperature, by a stiffening factor $\mu$
whose value is fixed by requiring that the H/D off-centering in the $xy$ plane, i.e., $\Delta_{xy}$ in Fig.~\ref{fig:crystal}(c), match the value estimated by neutron diffraction experiments~\cite{nelmes1987compilation, tibballs1982crystal, tibballs1982pt, nelmes1988onthestructural}. This procedure leads to $\mu=2.5$ and $\mu=16$, respectively, for H and D ions (see Methods). This empirical fix leaves the DFT potential energy surface unaffected, but lowers the zero point energy of H/D to compensate for the underestimation of the double well barrier in the H-bonds, enhancing the weight of the asymmetric \oho configurations. It turns out that static position-dependent observables are predicted with remarkable accuracy with this procedure. However, momentum-dependent observables, like the momentum distribution measured in deep inelastic neutron scattering experiments~\cite{reiter2002direct, reiter2008deuteron} are compromised and should not be calculated with this approach.

Using the above PIMD scheme we estimate the phase transition temperature $T_c$ from the temperature evolution of the spontaneous polarization $|\mathcal{P}_z|$, reported with the corresponding experimental data in Fig.~\ref{fig:bonds}(a). The estimated $T_c$ is $\approx120$K for KDP, and $\approx230$K for DKDP, with an error relative to experiments of less than 10K in both cases. The predicted saturated polarizations, for $T<T_c-20$K, closely match their experimental counterparts. Notably, the experimental observation that in KDP $|\mathcal{P}_z|$ is $\approx25\%$ smaller than in DKDP, is reproduced by the simulations, further supporting the validity of the model. Due to the finite size of the simulations, which include 512 KDP/DKDP units, the observed onset of spontaneous polarization is smoother than in experiments. A finite size analysis can be found in the Supplementary Information (SI), where we also report the calculated temperature dependence of the electric susceptibility, which shows approximate Curie-Weiss behavior near $T_c$ for both KDP and DKDP.

Experimentally, the PT in KDP/DKDP is first-order~\cite{reese1969studies}, structurally associated with a shear distortion of the lattice in the ferroelectric phase. The corresponding tetragonal to orthorhombic transformation is illustrated by the changes of the $a$ and $b$ lattice constants in Fig.~\ref{fig:bonds}(b). The agreement of theory and experiment is excellent. The models reproduce the measured thermal expansion for $T>T_c$. For $T<T_c$, they slightly underestimate the orthorhombic distortion, while predicting a less sharp transition than in experiments due to finite-size effects. The shift of the $a$ and $b$ curves upon deuteration is well described. In the SI, we also analyze the isotope effects on the lattice constant $c$, the unit cell volume $\Omega$, and the H-bond length $d_{\mathrm{O}...\mathrm{H}}$. The simulation reproduces the Ubbelohde effect~\cite{robertson1939structure, ubbelohde1939isotope}, i.e., the elongation of the H-bonds following deuteration, a fine effect in the order of the hundredths of an angstrom.

\begin{figure*}[tbh]
    \centering
    \includegraphics[width=0.9\linewidth]{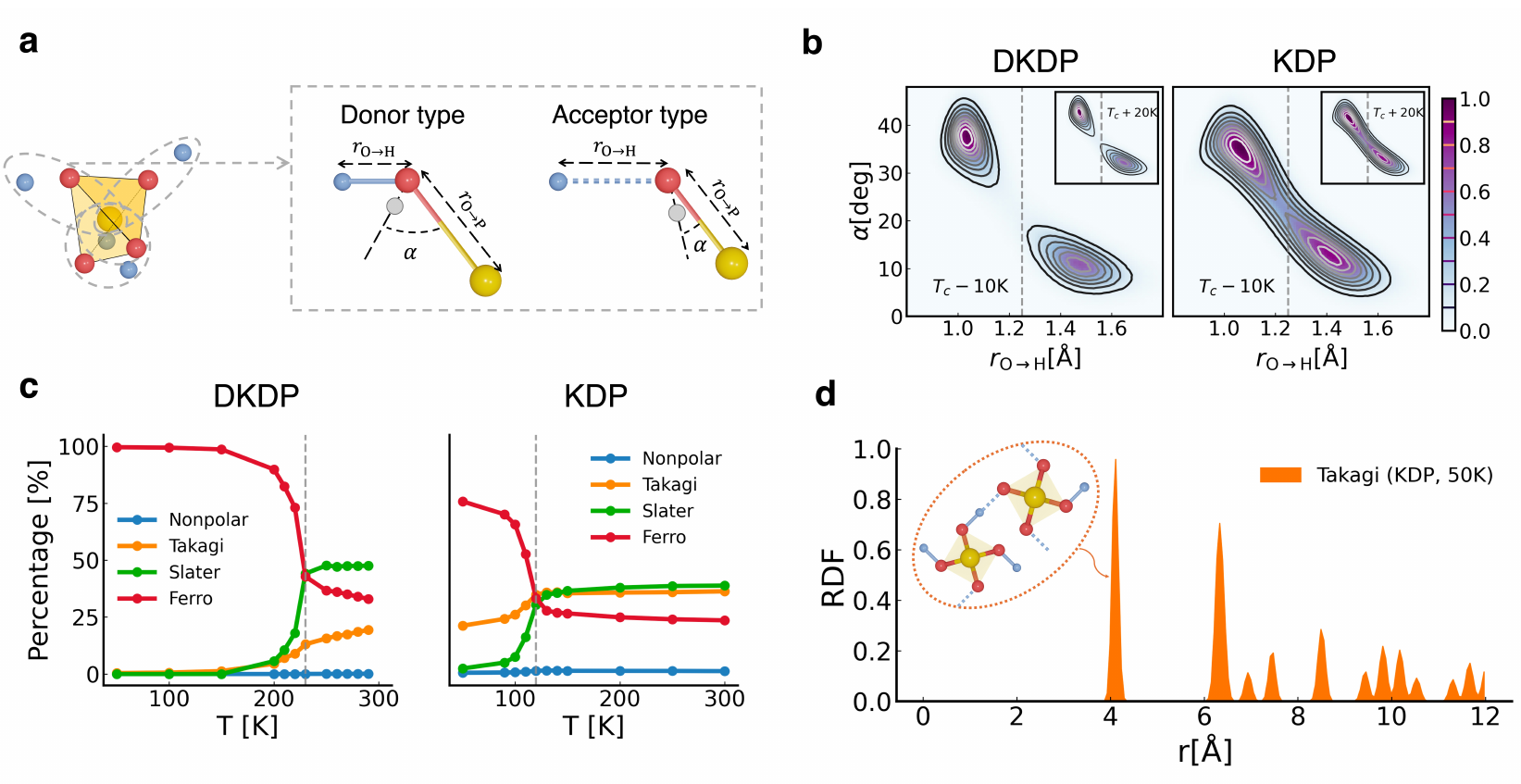}
    \caption{ 
    (a) Schematic representation of the rules for the classification of each KDP/DKDP replica. Each P-O-H (yellow-red-blue) structure can be donor (left) or acceptor (right) type. The grey sphere indicates
    the location of the electronic centroid.  
    (b) The joint probability distribution of $\alpha$ and $r_{\mathrm{O}\rightarrow\mathrm{H}}$ for KDP and DKDP, respectively, in the ferroelectric ($T=T_c-10$K) and the paraelectric phase (insets, $T=T_c+20$K). The vertical grey dashed line shows $r^*=1.25 \text{\AA}$. 
    (c) Percent population of the different classes of KDP/DKDP units vs temperature. The vertical grey dashed line shows $T_c$. (d) RDF of the Takagi defects in KDP at 50K. The distance $r$ between the defects is the spatial separation of the respective P ions.}
    \label{fig:defects}
\end{figure*}

The simulation gives access to microscopic properties that it would be difficult, if not impossible, to observe in experiments, such as the probability density function (PDF) of the local dipole moments $\mathbf{d}=(d_x, d_y, d_z)$.
The distributions $n(d_x,d_z)\equiv n(d_x,0,d_z)$ and 
$n(d_x,d_y)\equiv n(d_x,d_y,0)$ at two different temperatures are displayed in Fig.~\ref{fig:bonds}(c) for DKDP and KDP, respectively. In DKDP, breaking of the $z$-reflection symmetry of $n(d_x,d_z)$ for $T<T_c$ is evident in the upper $\mathrm{c_1}$ panel. This symmetry is restored for $T>T_c$ (upper $\mathrm{c_2}$ panel). The magnitude of $d_z$ in the peak of the distribution in panel $\mathrm{c_1}$ closely matches the dipole of the Ferro units (Fig.~\ref{fig:crystal}(d)). The peak magnitude of $d_z$ does not change appreciably across the transition ($\mathrm{c_2}$ panel), as expected for a PT dominated by order-disorder. In the lower 
$\mathrm{c_1}$ panel, $n(d_x,d_y)$ is uniformly close to zero, reflecting the $z$ alignment of the dipoles in the ferroelectric phase. In the paraelectric phase (lower $\mathrm{c_2}$ panel), $n(d_x,d_y)$ shows eight off-center spots with octagonal symmetry,
associated to the Slater units of Fig.~\ref{fig:crystal}(d). Each one of the four Slater units in that figure contributes to two off-center spots due to the glide reflection symmetry of the crystal.
The central peak in the lower $\mathrm{c_2}$ panel originates from the diffusive part of the two dominant peaks of the upper $\mathrm{c_2}$ panel.
The dipole distributions of KDP are reported in panels $\mathrm{c_3}$ and $\mathrm{c_4}$. 
Spontaneous symmetry breaking is again evident in $n(d_x,d_z)$, with a dipole magnitude that does not change much across the transition, consistent with order-disorder character. Relative to DKDP, the KDP distributions have a more strongly diffusive nature, signaling the presence of significantly more disorder above and below $T_c$. This is particularly evident in the paraelectric $n(d_x,d_y)$ distribution, which, in KDP, shows roughly circular symmetry
(lower $\mathrm{c_4}$ panel). As we will see, this is due to the prominent role played by Takagi units generated by quantum fluctuations. While in DKDP Slater units dominate the paraelectric phase near $T_c$, in KDP Takagi and Slater units are equally important, blurring the octagonal pattern of the lower $\mathrm{c_2}$ panel.

In PIMD simulations each configuration of the beads at equal imaginary time defines a distinct replica of the system (see Methods). Each KDP/DKDP unit in a replica can be assigned to one of the four classes displayed in     
Fig.~\ref{fig:crystal}(d), depending on the character of its four OH bonds, each one of which can have donor or acceptor H-bond character. The length and character of each OH bond is linearly correlated, to excellent approximation, with the length and character of the OP bond and the position of the electronic centroid about the oxygen (see Methods), as illustrated in Fig.~\ref{fig:defects}(a). It is because of this correlation that the dipole moment of a phosphate unit depends on the class to which it belongs. In view of this correlation, the two types of P-O-H(D) structure of Fig.~\ref{fig:defects}(a) can be distinguished by the length of $r_{\mathrm{O}\rightarrow\mathrm{H}}$.
In the following, we choose the value $r^* = 1.25\text{\AA}$ for the divider between donor and acceptor type structures for both KDP and DKDP below and above $T_c$. This choice is supported by the calculated joint PDF of $\alpha$ and $r_{\mathrm{O}\rightarrow\mathrm{H}}$,
i.e., $n(\alpha, r_{\mathrm{O}\rightarrow\mathrm{H}})$,
reported in Fig.~\ref{fig:defects}(b) for DKDP and KDP at $T=T_c-10$K, and at $T=T_c+20$K in the insets.
In all cases,  $n(\alpha, r_{\mathrm{O}\rightarrow\mathrm{H}})$ is strongly bimodal with the peaks corresponding to typical donor and acceptor type P-O-H(D) structures, respectively. Moreover, $n(\alpha, r_{\mathrm{O}\rightarrow\mathrm{H}})$
changes little across the PT.
In DKDP, the divider $r^*$, indicated by the grey dashed line in the figure, separates well donor and acceptor configurations. In KDP, quantum fluctuations make donor and acceptor distributions overlap, but symmetric H-bond configurations, in which H lies approximately in the middle of the O-O bond, have a substantially smaller statistical weight than asymmetric configurations, even at the lowest temperature of the simulations ($T=50$K), indicating that our choice of $r^*$ is meaningful.

The temperature variation of the population of phosphate units, classified using $r^*$, is plotted in Fig.~\ref{fig:defects}(c). For $T<T_c$, broken symmetry Ferro units dominate both DKDP and KDP. Long-range order is lost in the paraelectric phase where the Ferro units, while still present, are not dominant and the majority of the units are Slater and/or Takagi. In this phase, the frequent interconversions of Ferro, Slater and Takagi units, restore the $z\rightarrow -z$ symmetry of the Ferro units. The population of Nonpolar units is negligible at all temperatures. The most important message of Fig.~\ref{fig:defects}(c)
is that Slater and Takagi populations are markedly different in DKDP and KDP. In DKDP, when $T<150$K almost all units are Ferro.  Dipolar defects, mostly associated with Slater units, appear only for $T>150$K, causing a gradual reduction of the spontaneous polarization. By contrast, in KDP, a substantial fraction of Takagi, rather than the more stable Slater units, persists down to $T=50K$, whereby Ferro units only account for $\approx 75\%$ of the phosphates. On the time scale of the PIMD simulations, the Takagi units are dynamic and fluctuate with H-bond switches every several picoseconds. By contrast, H-bond switches require tens of picoseconds at low temperatures in DKDP.  Given that the molecular dipole of an ideal Ferro unit is only marginally affected by deuteration, the roughly $25\%$ smaller polarization of KDP relative to DKDP can only originate from the dipolar defects present in KDP, and is mostly absent in DKDP at low temperatures. Dipolar disorder also explains why in experiments
the excess transition entropy is $\approx30\%$ smaller in KDP than in DKDP~\cite{strukov1968comparative}. 

The Takagi defects are spatially correlated as shown by their radial distribution function (RDF) in KDP at $T=50$K in Fig.~\ref{fig:defects}(d). Due to averaging over the replicas in imaginary time, this RDF is already well converged in a single PIMD snapshot. The first peak corresponds to pairs of adjacent defects created mostly by switches of the H-bonds between neighboring Ferro units. The weak temperature dependence of the Takagi fraction in Fig.~\ref{fig:defects}(c) suggests that the effect should originate from the quantum tunneling of the protons observed in deep inelastic neutron scattering~\cite{reiter2002direct}. Interestingly, the same experiments do not provide tunneling evidence in DKDP~\cite{reiter2008deuteron}. The prominent second peak of the RDF indicates that an additional defect is likely to reside near a defect pair, while the non-monotonic decay of the RDF with the Euclidean distance reflects a monotonic decay with the graph distance defined as the minimal number of \oho bridges connecting pairs of Takagi defects. The defects are also dynamically correlated as suggested by analyses of their joint PDF reported in the SI. Recombination and migration to a neighboring site are two sources of dynamical correlation. The Takagi defects are nominally charged and correlation acts to reduce local charge fluctuations.      

 
 
Because of quantum disorder, the ferroelectric state of KDP is fundamentally different from that of DKDP. Linear extrapolation in Fig.~\ref{fig:defects}(c) suggests that a substantial fraction ($\approx15\%$) of fluctuating Takagi defects should be present near absolute zero, providing a possible realization of the TLS that explains the anomalous heat capacity of insulating glasses near $T=0$K~\cite{anderson1972anomalous,phillips1987two}. This speculation is supported by the experimental findings of an anomalous heat capacity varying approximately linearly with $T$ near absolute zero in KDP~\cite{lawless1976t, lawless1982specific,foote1985low}. The TLS in KDP should be an intrinsic equilibrium property rather than the outcome of quenched glassy disorder, but it would be difficult to test this hypothesis as it is hard to control the various sources of extrinsic disorder present in the experimental samples~\cite{lawless1976t}. 

The huge effect of quantum fluctuations in 
KDP requires a fine balance between the tunneling barrier and the zero point energy. This balance, which is broken by deuteration with the ensuing large isotope effects, is realized in the KDP family of crystals due to the presence of unusually strong H-bonds. H-bonds can be made stronger by pressure. In water ice, where the H-bonds at ambient conditions are relatively strong but significantly weaker than in KDP, isotope effects are relatively small, but they become large at pressures in the hundreds of kbar range~\cite{pruzan1994pressure}. In KDP, experiments show that the temperature of the ferroelectric transition is reduced with increasing pressure, until, at pressure in excess of 17kbar the PT disappears and the system becomes a quantum paraelectric~\cite{samara1987pressure, Endo2002quantum}. We understand this phenomenology as the result of an increasing Takagi population following increasing H-bond symmetrization with pressure. We speculate that the phase point at which the PT  disappears ($T=0$K, $P=17$kbar) may be associated to quantum criticality~\cite{sachdev1999quantum, rowley2014ferroelectric}, an hypothesis consistent with the large dielectric susceptibility measured in the vicinity of this point~\cite{Endo2002quantum}. In future studies, it would be of interest to look for possible non-classical temperature dependence of the inverse dielectric susceptibility at very low temperature~\cite{rowley2014ferroelectric}.



\clearpage

\section*{Methods}\label{methods}

\subsection*{Density functional theory and atomistic models}

We adopt SCAN, the strongly constrained and appropriately normed functional approximation for exchange and correlation of Ref.~\cite{sun2015strongly}. It has
been reported that SCAN outperforms PBE and LDA in modeling the structural properties of KDP~\cite{zhang2017comparative}. We perform electronic structure calculations with Quantum Espresso~\cite{giannozzi2009quantum}, using norm-conserving pseudopotentials \cite{hamann2013optimized} that include in the valence the $2s$ and $2p$ states of O, the $3s$ and $3p$ states of P, and the $3s$, $3p$, and $4s$ states of K. 
We find a stable classical ferroelectric structure of space group Fdd2 at $T=0\text{K}$ (Fig.1(a)), with lattice constants $a=10.62 \text{\AA}$, $b=10.67 \text{\AA}$, and $c=6.86 \text{\AA}$. The length of the O-H covalent bond is $1.042 \text{\AA}$, that of the O...H H-bond is $1.438 \text{\AA}$, and the H off-centering is equal to $\Delta_{xy}^{\text{Fdd2}}=0.198\text{\AA}$. The above results agree well with Ref.~\cite{zhang2017comparative} with minor differences due to the different pseudopotentials and numerical methods adopted in the two calculations.
  
We use DeePMD-Kit~\cite{wang2018deepmd} to train a DP model~\cite{zhang2018deep} for the potential energy surface within SCAN-DFT. The typical error of the model relative to DFT is less than $1 \text{meV/atom}$. Training details and error distributions are reported in the SI. 
In addition, we train a Deep Wannier (DW) model~\cite{zhang2020deepwannier} for the polarization surface in terms of the Wannier decomposition of the valence electronic structure~\cite{resta2007theory}.  
Maximally localized Wannier functions are computed with Wannier90~\cite{Pizzi2020} for
a subset of the atomic configurations in the DP training set (see SI). Each Wannier distribution accommodates two spin-degenerate electrons. In all the atomic configurations four Wannier centers are uniquely associated to each O and four others to each K atom, forming approximate atom-centered tetrahedral structures. The Wannier centroids (WCs), i.e., the geometric centers of the four Wannier centers associated to O or K atoms provide all the necessary information to compute the polarization~\cite{zhang2020deepwannier}. The DW model gives the environmental dependence of the WCs. While the oxygen WCs depend crucially on the chemical environment, as illustrated in Fig.~\ref{fig:defects}(a), the potassium WCs, associated with the $3s$ and $3p$ semi-core states, are essentially independent of the environment and their contribution to polarization changes is negligible. We do not include them in the DW model, i.e., to calculate the polarization we treat the semi-core states as frozen core electrons of the $\mathrm{K^{+}}$ ion. The training of the DW model is done with DeePMD-Kit. The details are reported in the SI.

Following the theory of polarization~\cite{resta2007theory, vanderbilt2018berry},
we assign to each elementary KDP/DKDP unit ( Fig.~\ref{fig:crystal}(b)) a molecular dipole $\mathbf{d}$ given by:
\begin{equation}\label{molecular dipole}
\begin{split}
\mathbf{d} = \sum_{i=1}^{2} q_\mathrm{K} (\mathbf{r}_\mathrm{K}^{(i)}-\mathbf{r}_\mathrm{P}) + \sum_{i=1}^{4} q_\mathrm{O} (\mathbf{r}_\mathrm{O}^{(i)}-\mathbf{r}_\mathrm{P}) \\
+ \sum_{i=1}^{4} q_\mathrm{H} (\mathbf{r}_\mathrm{H}^{(i)}-\mathbf{r}_\mathrm{P}) + 
 \sum_{i=1}^{4} q_\mathrm{WC} (\mathbf{r}_\mathrm{WC}^{(i)}-\mathbf{r}_\mathrm{P}).
\end{split}
\end{equation} 
Here, $\mathbf{r}_\mathrm{K}^{(i)}$, $\mathbf{r}_\mathrm{O}^{(i)}$, $\mathbf{r}_\mathrm{H}^{(i)}$, and $\mathbf{r}_\mathrm{P}$ are the position vectors of the K, O, H/D, and P ions belonging to the elementary unit, and $q_\mathrm{K}=0.5e$, $q_\mathrm{O}=6e$, and $q_\mathrm{H}=0.5e$ are the charges carried by the K, O, and H/D ions, respectively. Since the K and H/D ions are shared with a neighboring unit, they carry half charge. The P ion, carrying a charge of $5e$, is taken as the local reference and does not contribute to the dipole. Finally,  
$\mathbf{r}_\mathrm{WC}^{(i)}$ are the position vectors of the WCs associated with the O ions and $q_\mathrm{WC}=-8e$ are their charges. When all the ionic and electronic charges are taken into account, each elementary unit is electrically neutral. The positions of the ions and of the WCs corresponding to each atomic configuration $\mathbf{R}=\{\mathbf{r}_1,\cdots,\mathbf{r}_N \}$ visited by PIMD in a periodic simulation box of volume $V$ and $N$ physical particles are provided by the DP and DW models. The global dipole of the configuration $\mathbf{R}$ is $\mathbf{D}(\mathbf{R}) = \sum_j \mathbf{d}_j$, where the sum extends to all the local molecular dipoles $\mathbf{d}_j$ contained in the simulation box. The polarization surface $\boldsymbol{ \mathcal{P}}(\mathbf{R})$, i.e., the polarization of the configuration $\mathbf{R}$,
is given, modulo an immaterial constant, by 
$\boldsymbol{ \mathcal{P}}(\mathbf{R})=\mathbf{D}(\mathbf{R})/V$~\cite{resta2007theory}. 

\subsection*{Path integral molecular dynamics}

Within Feynman's path integral formulation of quantum statistical mechanics~\cite{feynman2010quantum}, the canonical partition function of $N$ distinguishable quantum particles can be approximated with a Trotter factorization in imaginary time:

\begin{equation}\label{partition}
Z_L  \approx \int \mathcal{D}\mathbf{R} e^{-\beta \mathcal{H}^{\text{eff}}_L},
\end{equation}
where
\begin{equation}
\mathcal{D}\mathbf{R} \equiv \prod_{i=1}^N\left(\frac{  m_i L}{2 \pi \beta\hbar^2}\right) ^{\frac{3L}{2}} \text{d}\mathbf{r}_i^{(1)}\cdots \text{d}\mathbf{r}_i^{(L)}.
\end{equation}

The effective Hamiltonian $\mathcal{H}^{\text{eff}}_L$ is
\begin{equation}\label{hamiltonian}
\mathcal{H}^{\text{eff}}_L =\sum^L_{s=1}\Bigg(  \sum_{i=1}^N
    \frac{k_i}{2}\left(\mathbf{r}^{(s)}_i-\mathbf{r}^{(s+1)}_i\right)^2+\frac{1}{L} U({\mathbf R}^{(s)})\Bigg),
\end{equation}
where $k_i = \frac{m_iL}{\beta^2\hbar^2}$ and $\mathbf{R}^{(s)} = \{ \mathbf{r}_1^{(s)}, \cdots, \mathbf{r}_N^{(s)}\}$. The index $s$ labels imaginary times, the positive integer $L$ is the total number of imaginary time slices, and the condition
$L+1=1$ applies in the argument of the sum over the beads so that $\mathbf{r}^{(L+1)}_i=\mathbf{r}^{(1)}_i$.
Thus, a discretized Feynman path in imaginary time, $\{ \mathbf{r}_i^{(1)}, \cdots, \mathbf{r}_i^{(L)}\}$, is equivalent to a ring polymer. $U$ is the potential energy surface provided by the DP model. $Z_L$ recovers the exact quantum partition function in the limit $L\rightarrow \infty$. For finite $L$, Eq.~(\ref{partition}) can be seen as the configurational partition function of a classical system of $N$ ring polymers each made of $L$ beads and described by the effective Hamiltonian $\mathcal{H}^{\text{eff}}_L$, in which the constant
$k_i$ defines the spring that links neighboring beads in imaginary time, and $\frac{1}{L} U({\mathbf R}^{(s)})$ is the potential of interaction between different ring polymers at equal imaginary time. The ensemble of all the beads at equal imaginary time constitutes a replica of the system. PIMD~\cite{tuckerman2010statistical} is a molecular dynamics scheme for sampling the configurations of the ring polymers with Boltzmann weights proportional to $e^{-\beta \mathcal{H}^{\text{eff}}_{L}}$. Then, a canonical average of a position-dependent observable of the quantum system is estimated from a classical canonical average of the ring-polymer system. For example, the average macroscopic polarization is calculated as follows: 
\begin{equation}
 \begin{split}
   \boldsymbol{ \mathcal{P}} = \frac{1}{L}\left\langle \sum_{s=1}^L \boldsymbol{ \mathcal{P}}(\mathbf{R}^{(s)})
   \right\rangle_{Z_L} =             \\
   \frac{\int \mathcal{D}\mathbf{R}  \sum_{s=1}^L \boldsymbol{ \mathcal{P}}(\mathbf{R}^{(s)}) e^{-\beta \mathcal{H}^{\text{eff}}_L}}{ L \int \mathcal{D}\mathbf{R}  e^{-\beta \mathcal{H}^{\text{eff}}_L }},
 \end{split}  
\end{equation}
where $\boldsymbol{ \mathcal{P}}(\mathbf{R}^{(s)})$, the polarization of the $s$-th replica, is provided by the DW model.
 
\subsection*{PIMD simulations}

We use I-PI~\cite{kapil2019pi}, LAMMPS~\cite{thompson2022lammps}, and DeePMD-Kit to perform NPT-PIMD simulations on a supercell  
with 512 KDP/DKDP units in the temperature range $[50, 300]\text{K}$. The temperature is controlled with the PILE thermostat \cite{ceriotti2010efficient} and the pressure is maintained with an anisotropic barostat~\cite{Martyna1999}. The spring constants $k_i$ (Eq.~(\ref{hamiltonian})) associated with K, P, and O ions take their physical values, but the $k_i$ associated with H is multiplied by a stiffening factor $\mu=2.5$, while the one corresponding to D is multiplied by $\mu=16$. These values are chosen to fit the experimental neutron data on the off-center displacement of H/D, as explained in the next section.

To control convergence with $L$, we use the estimator $K_{L}$ of the thermodynamic quantum kinetic energy per particle suggested in Ref.~\cite{ceriotti2012efficient}:
\begin{equation}
K_{L} = \frac{3}{2\beta} + \frac{1}{2LN} \Bigg\langle\sum_{s=1}^L \sum_{i=1}^N
\left({\mathbf r_i^{(s)}} - \overline{ {\mathbf r_i^{}} } \right)
\cdot \frac{\partial U}{\partial {\mathbf r_i^{(s)}} } \Bigg\rangle_{Z_L}.
\end{equation}
Here, $\overline{ {\mathbf r_i^{}}}$ is the centroid of all the beads of particle $i$. In the classical limit, $L=1$ and $K_L = 3k_BT/2$, the classical kinetic energy per particle. By monitoring the convergence of $K_L$ with $L$ (see SI) we estimate $|K_L-K_\infty|$ at different temperatures. In the production runs we adopt $L$ values for which $|K_L-K_\infty|$ is smaller than $1 \text{meV/atom}$, which is also the approximate error of the DP representation. A study of the effect of finite $N$ is impractical with PIMD simulations as $L$ can be as large as $128$. For that reason, we only report in the SI a study of the effect of $N$ on the corresponding classical system ($L=1$).  
\begin{figure}[hbt]
    \centering
    \includegraphics[width=0.85\linewidth]{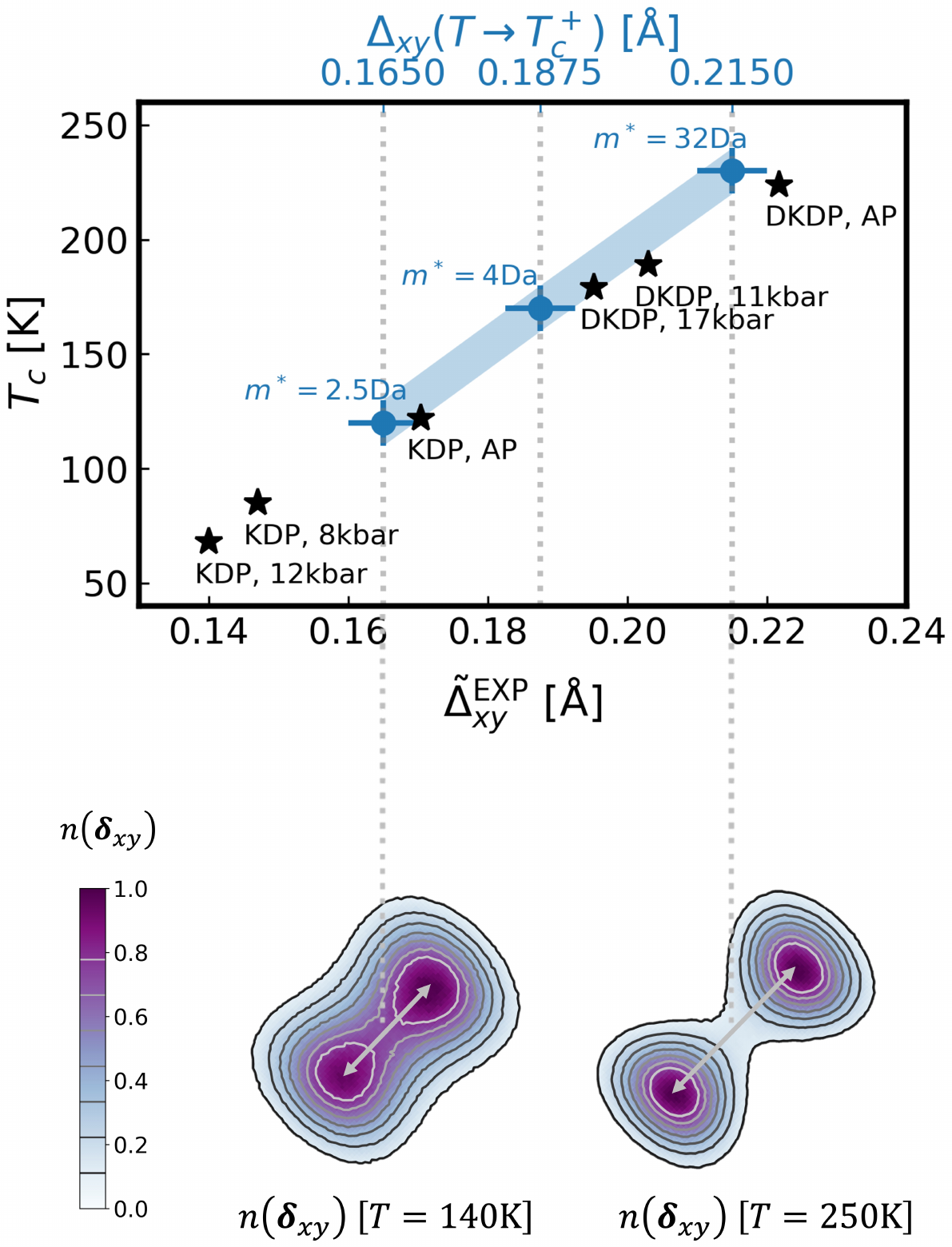}
    \caption{(Upper) The relation between $T_c$ and H/D off-centering. The black stars mark the experimental measurements of $T_c$ and $\tilde{\Delta}_{xy}^{\mathrm{EXP}}$ for KDP and DKDP at different external pressures~\cite{nelmes1987compilation, tibballs1982crystal, tibballs1982pt, nelmes1988onthestructural}. Blue spheres mark the numerical $T_c$ and $\Delta_{xy}(T\rightarrow T_c^+)$ obtained with PIMD with different $m^*$ at atmospheric pressure. The error bar of $T_c$ is $\pm 10$K, reflecting the $T$
    spacing adopted in the calculations of the order parameter near $T_c$. The error bar of $\Delta_{xy}(T\rightarrow T_c^+)$ is $\pm 0.005\text{\AA}$, reflecting the space resolution of the histogram representing $n(\boldsymbol \delta_{xy})$ from which we extract $\Delta_{xy}(T\rightarrow T_c^+)$. Blue shadows are a guide for the eye, suggesting an almost linear $T_c$-$\Delta_{xy}(T\rightarrow T_c^+)$ relation. 
    (Lower) The heatmap of $n(\boldsymbol \delta_{xy})$ computed at $T>T_c$. The left figure is for $m^* =2.5$Da and $T=150$K. The right figure is for $m^*=32$Da and $T=250$K.
    }
    \label{fig:Tc}
\end{figure}

\subsection*{The spring stiffening factor $\mu$ for H and D ions}


In KDP, SCAN-DFT predicts a ferroelectrically ordered classical ground state~\cite{PhysRevB.52.6301}, but this structure is destabilized by NQE. Absence of ferroelectricity was also found in ab initio PIMD simulations of KDP with the PBE-DFT functional~\cite{srinivasan2011isotope}. This occurs because common functionals tend to underestimate the double-well potential energy barrier in the H-bonds, wiping out the classical ferroelectric ground state~\cite{menchon2018ab}. 

In the presence of NQE, a correct description of the ferroelectric phase requires a delicate balance of tunneling barrier and zero-point energy of the protons, which are both strongly correlated with the geometric distortion of the phosphate groups, as shown in numerous experimental~\cite{nelmes1987compilation,  tibballs1982crystal, tibballs1982pt, nelmes1988onthestructural, mcmahon1990geometric} and theoretical~\cite{ichikawa2001structural, koval2002ferroelectricity,koval2005first} studies. 

Experiments find that, in KDP and DKDP, $T_c$ at different pressure conditions is to very good approximation linearly correlated with $\tilde{\Delta}_{xy}^{\mathrm{EXP}}$, the off-center displacement of the H/D ions in the paraelectric phase just above the PT
~\cite{nelmes1987compilation, tibballs1982crystal, tibballs1982pt, nelmes1988onthestructural}. This correlation, shown in Fig.~\ref{fig:Tc}, strongly supports the notion that H/D off-centering is the key geometric parameter controlling $T_c$, and hints at a possible way of improving the experimental agreement of DFT-based PIMD simulations by fixing the predicted H/D off-centering. In KDP, SCAN-based PIMD always yields unimodal proton distributions with $\Delta_{xy}(T)=0$, but the excessive quantum delocalization of H/D, responsible for the effect, can be reduced as desired by multiplying the H/D spring constant $k = \frac{m L}{\beta^2\hbar^2}$ by a stiffening factor $\mu$ greater than 1, a procedure equivalent to assigning an effective mass $m^*=\mu m$ to H/D. The displacements $\Delta_{xy}(T\rightarrow T_c^+)$ and the corresponding $T_c$ calculated with PIMD for three different values of $m^*$ at ambient pressure (AP) are reported in  Fig.~\ref{fig:Tc}, showing a linear behavior of $\Delta_{xy}$ with $T_c$ that closely mimics the experimental one. For $m^*=2.5\text{Da}$, i.e. $\mu=2.5$, and for $m^*=32\text{Da}$, i.e., $\mu=16$, $\Delta_{xy}(T\rightarrow T_c^+)$ and $T_c$ match the experimental displacements and transition temperatures of KDP and DKDP, respectively. This is the choice of $\mu$ that we adopt in our PIMD simulations of KDP and DKDP. 

The displacement $\Delta_{xy}(T\rightarrow T_c^+)$ is estimated from thermal averages at $T\approx T_c+20$K to reduce the large impact of the fluctuations near $T_c$. The probability distribution functions $n(\boldsymbol \delta_{xy})$ for H and D are also reported in Fig.~\ref{fig:Tc}, showing a bimodal character in qualitative agreement with neutron diffraction experiments~\cite{nelmes1987structural}. Relative to KDP, DKDP has a more localized $n(\boldsymbol \delta_{xy})$ and a larger $\Delta_{xy}$, consistent with the more classical behavior of the deuterated H-bonds. In fact, in the classical limit, when all the atoms in PIMD have infinite mass, $T_c$ is approximately equal to $250$K, slightly above the predicted transition temperature of DKDP, consistent with a predominantly classical character of the transition in that system. 


\section*{Acknowledgements}
We thank Yifan Li, Chunyi Zhang, Xifan Wu, 
Chenyuan Li, Chenxing Luo, and Linfeng Zhang for fruitful discussions. All authors were supported by the Computational Chemical Sciences Center: Chemistry in Solution and at Interfaces (CSI) funded by DOE Award DE-SC0019394. The authors are pleased to acknowledge that the work reported in this paper was performed largely using the Princeton Research Computing resources at Princeton University. This research also used resources of the National Energy Research Scientific Computing Center (NERSC) operated under Contract No. DE-AC02-05CH11231 using NERSC award ERCAP0021510.


\bibliographystyle{naturemag}
\bibliography{references.bib}

\end{document}


\title{Supplementary Information}
\maketitle
 
\section{Details of the Deep Potential and Deep Wannier models}\label{models}

\begin{figure}[h!]
\centering\includegraphics[width=\linewidth]{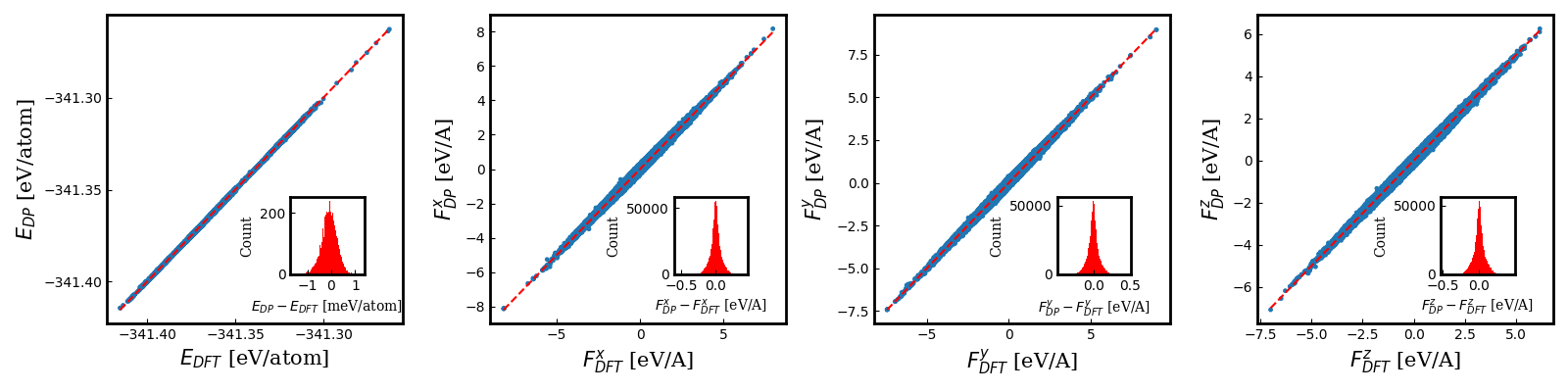}
    \caption{Comparison of energies and forces predicted by the DP model with the reference SCAN-DFT results in the training dataset. The insets show the error distribution.}
    \label{fig:dp}
\end{figure}

We used the active learning agent DPGEN~\cite{zhang2020dp} to collect DFT data in the temperature range $[50, 300]$K and the pressure range $[-20, 100] $kbar. The active-learning exploration of atomic configurations is performed for 23 iterations. 
The final training dataset contains 5800 atomic configurations. The performance of the DP model on the training set is shown in Fig.~\ref{fig:dp}. The test set is collected from PIMD simulations at $T=100$K and  $T=150$K. The prediction error is smaller than 1meV/atom in the potential energy and smaller than 0.25eV/$\text{\AA}$ in the Cartesian components of the forces.
The error levels in the training and test sets are similar. 

\begin{figure}[h!]
    \centering
\includegraphics[width=0.7\linewidth]{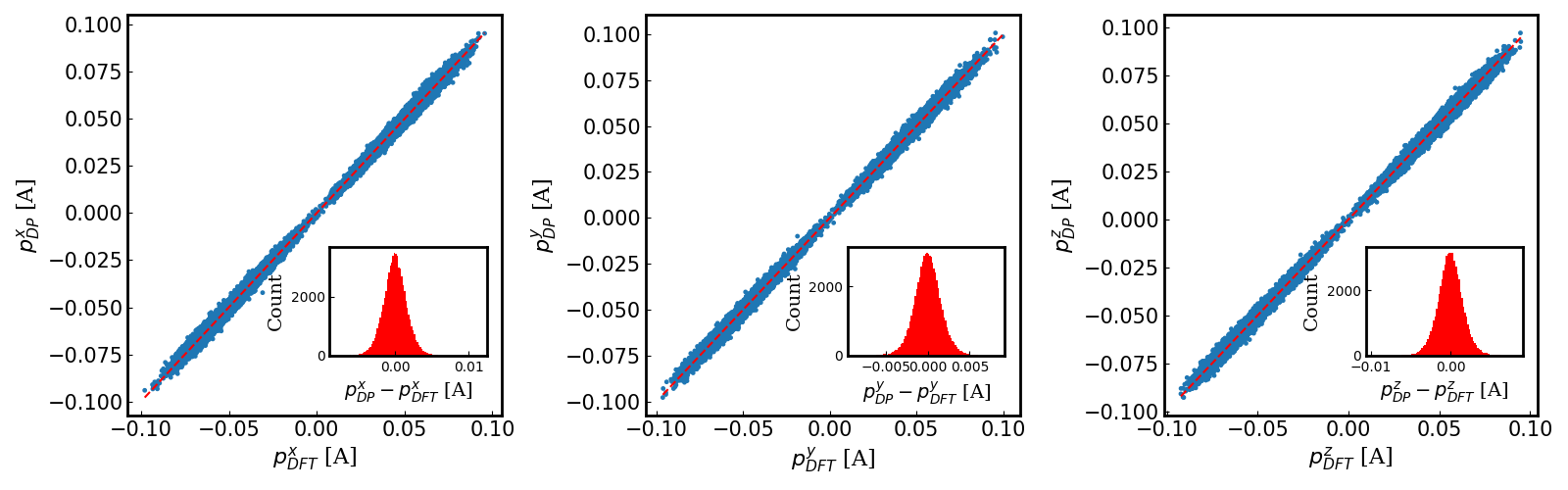}
    \caption{Comparison of the displacement vectors of the Wannier centroids from their associated O atom predicted using the DW model with the reference DFT results in the training dataset. The insets show the error distribution.}
    \label{fig:dipole}
\end{figure}

Then, from the atomic configurations generated through the first 10 iterations of active-learning exploration, we randomly collect 100 atomic configurations from each iteration. The 1000 collected configurations form a reduced dataset for the DW model. The reduction of the dataset is due to the high computational cost of generating training labels for the DW model, and the fact that the full dataset generated by active learning is typically redundant. 
The maximally localized Wannier functions corresponding to these atomic configurations are calculated with Wannier90~\cite{Pizzi2020}. The training labels for the DW model are the displacement vectors from the oxygens of the associated Wannier centroids. As shown in Fig.~\ref{fig:dipole}, the standard error of the DW model is 0.002$\text{\AA}$ for the Cartesian components of the displacement vectors, much smaller than the typical distance from the oxygen of the associated Wannier centroid.

\section{Convergence of PIMD with the number of beads}\label{beads}

\begin{figure}[h!]
    \centering
\includegraphics[width=0.7\linewidth]{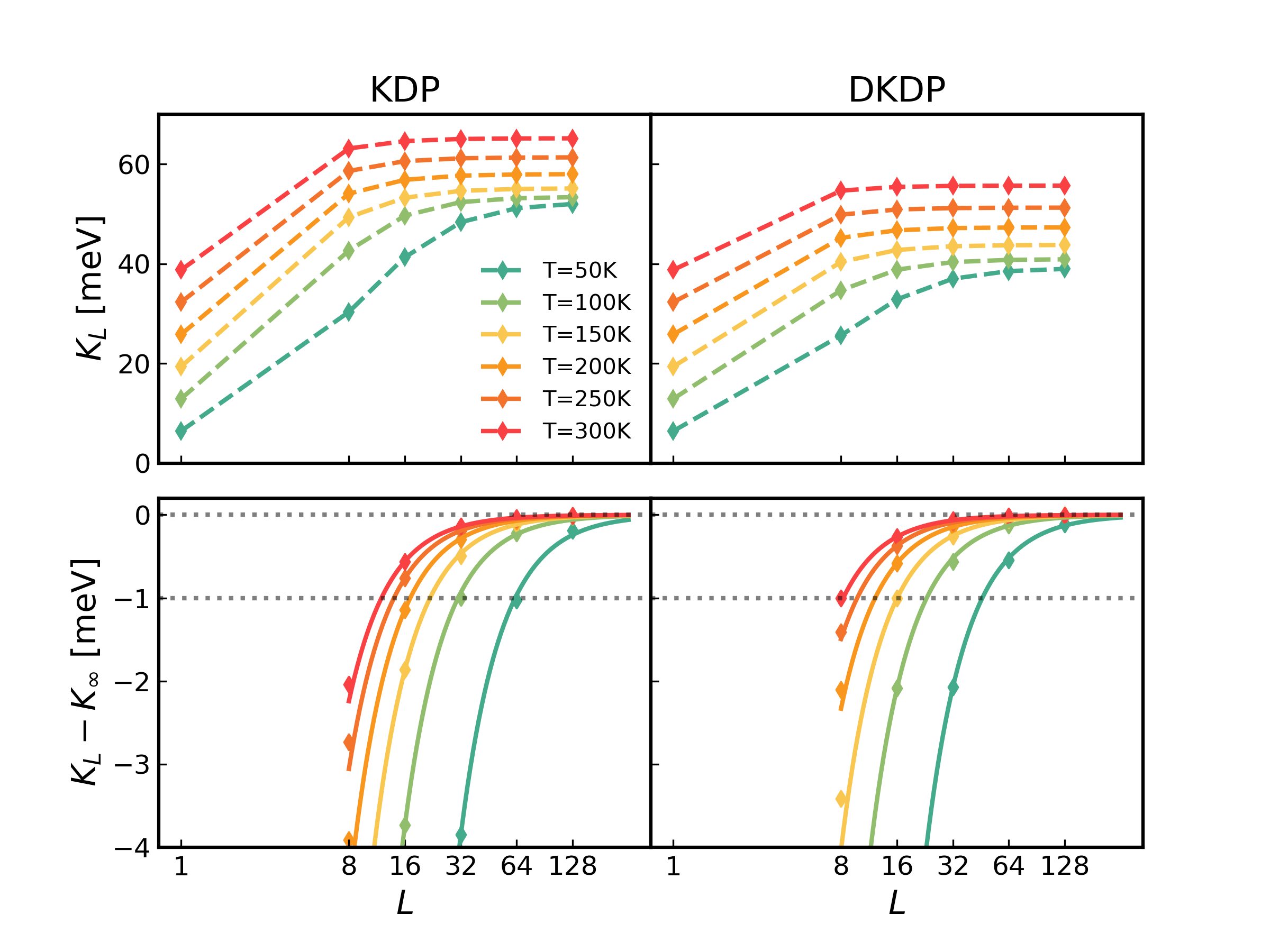}
    \caption{(Upper) The quantum kinetic energy $K_L$ as a function of the number of beads. (Lower) $K_L-K_\infty$ as a function of the number of beads. The diamonds are the numerical results from PIMD. The solid lines are the curves from optimal fitting. The two horizontal dotted guiding lines are at $K_L-K_\infty=0$ and $K_L-K_\infty=-1$meV, respectively. }
    \label{fig:beads}
\end{figure}

To fix $L$, the number of beads needed for good convergence of quantum statistical averages, we study the convergence with $L$ of the thermodynamic quantum kinetic energy $K_L$. We perform PIMD simulations of systems with 512 KDP/DKDP units at different temperatures ($T=50,100,150,200,250,300$K) with the adopted stiffening factor for H and D ions, respectively. The calculated $K_L$ as a function of $L$ is shown in the upper panel of Fig.\ref{fig:beads}. $K_L$ converges more slowly at lower temperatures. About $128$ beads are necessary at $T=50$K, the lowest temperature probed in our simulations. 
To estimate the error of a finite $L$ calculation, we estimate $K_\infty = \lim_{L\rightarrow\infty }K_L$, by fitting the $K_L$ data with the function $K_L=K_\infty - h/L^2$ suggested by the $L$ dependence of the Trotter factorization error for a path of length $L$. We employ a non-linear least square curve fitting using the Python package SciPy~\cite{2020SciPy-NMeth}, treating $K_\infty$ and $h$ as parameters. At $T=50$K, $K_L$ data with $L\geq 32$ are used for fitting. At all the other temperatures, $K_L$ data with $L\geq 16$ are used. The fitted curves for $K_L-K_\infty$ are displayed in the lower panel of Fig.\ref{fig:beads}. We consider a calculation converged with $L$ when the error is less then 1meV/atom, which is close to the typical error of our DP model. With this criterion we find that in KDP at $T\geq 150$K one needs $L=32$. When $150$K$>T\geq 100$ K, $L$ is set to $64$, while for $100$K$>T\geq 50K$,  $L$ is set to $128$. In DKDP, when $T>200$K we set  
$L\geq 16$, and we set $L$ to 
$L=64$ when $T\leq 200$K.

\section{Classical finite size effects}\label{classical}

We study finite size effects on the ferroelectric PT of KDP with classical MD, i.e., by setting $L=1$ in PIMD.  
 \begin{figure*}[tbh]
    \centering
    \includegraphics[width=0.6\linewidth]{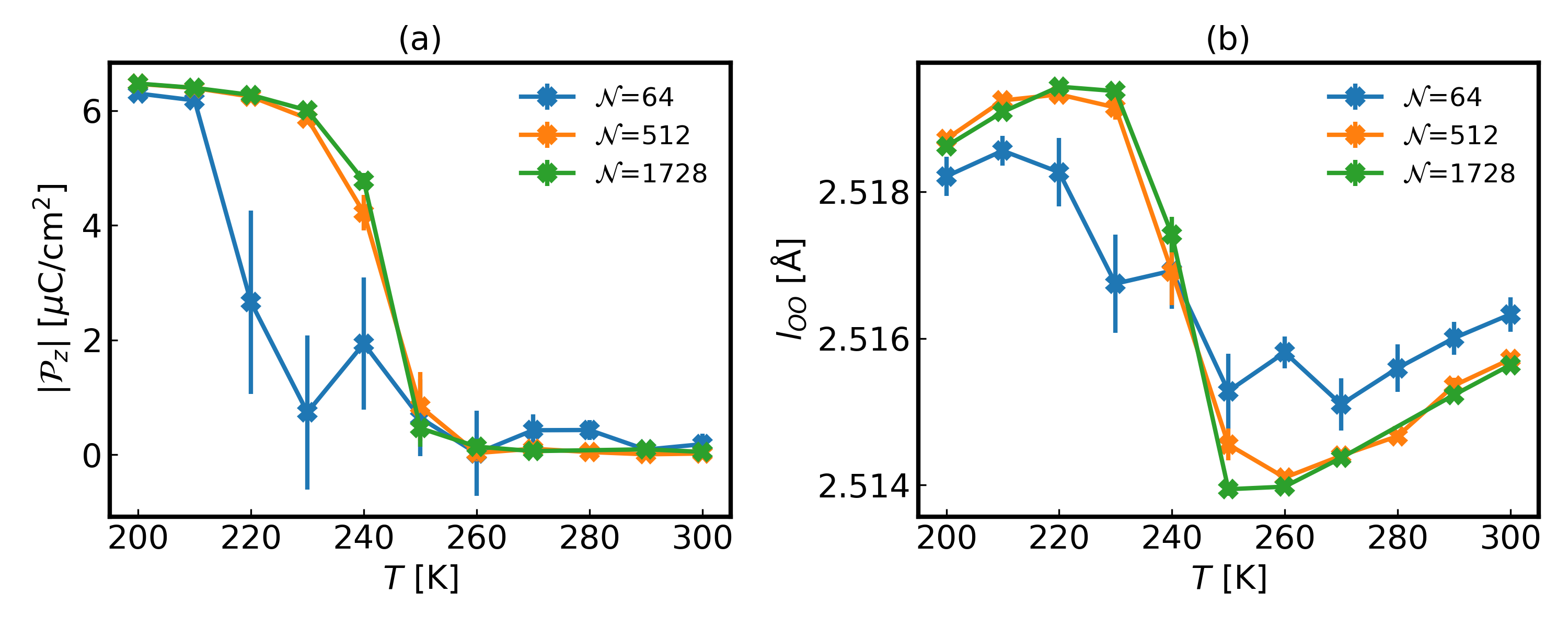}
    \caption{(a) $|\mathcal{P}_z|$ as a function of temperature from classical MD. (b) Average $l_{\text{OO}}$ as a function of temperature from classical MD.}
    \label{fig:order}
\end{figure*}
Calling $\mathcal{N}$ the number of KDP units in the simulation supercell, we compute $500 \text{ps}$ long classical NPT trajectories  for $\mathcal{N}=64$, $512$, and $1728$ at different temperatures and standard pressure conditions. The finite size effect on the PT is illustrated by the temperature dependence of the polarization in Fig.\ref{fig:order}(a) and by the temperature dependence of the average nearest neighbor O-O distance in Fig.\ref{fig:order}(b). The figure suggests that the size effect is small when    
$\mathcal{N}\geq512$. A minor sharpening of the transition onset can be seen as $\mathcal{N}$ increases from $512$ to $1728$, suggesting that significantly larger sizes would be needed to reproduce more closely the sharp onset observed in experiment. From the transition onset in Fig.\ref{fig:order} we estimate that $T_c$ should be in the range $[245, 255]$K. 
The estimated change of $l_{OO}$ upon the phase transition is approximately $0.005 \text{\AA}$, close to experimental findings. 
We expect that similar size effects should affect the PIMD simulations with $L\geq16$ reported in the manuscript.  

\section{Geometric and dielectric isotope effects} 

\begin{figure}[h!]
    \centering
    \includegraphics[width=\linewidth]{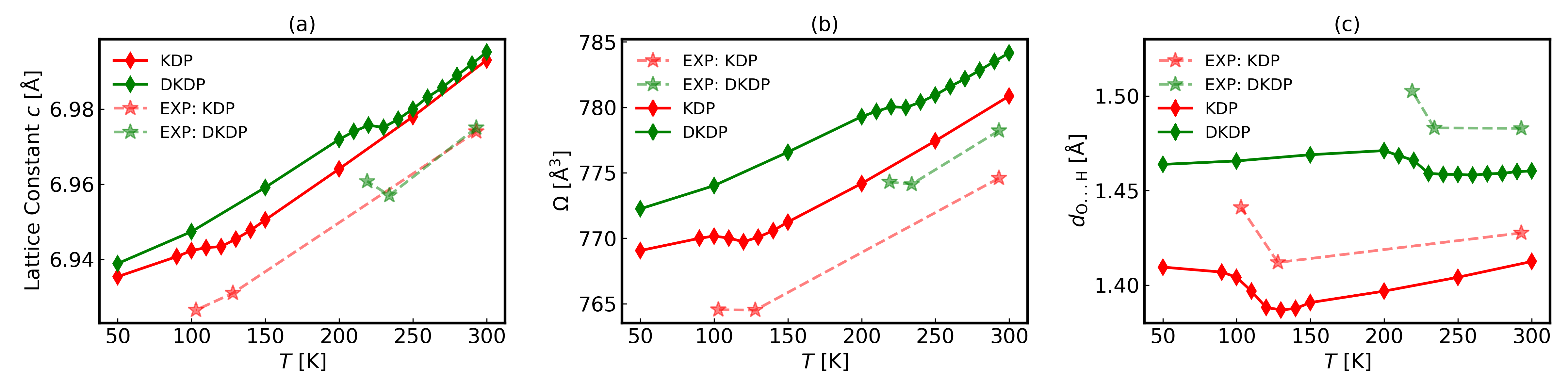}
    \caption{ (a) Predicted and experimental~\cite{nelmes1987compilation}  lattice constant $c$ as a function of $T$. (b) Predicted and experimental~\cite{nelmes1987compilation}
    unit cell volume $\Omega=abc$ as a function of $T$. (c) Predicted and experimental~\cite{nelmes1987compilation}
    H-bond length as a function of $T$.}
    \label{fig:latt}
\end{figure}

 The effects of phase transition and deuteration on the lattice constant $c$, the unit cell volume $\Omega$, and the H-bond length $d_{\mathrm{O}...\mathrm{H}}$ are illustrated in Fig.~\ref{fig:latt}, showing good agreement of theory and experiment. The Ubbelohde effect on $d_{\mathrm{O}...\mathrm{H}}$ ~\cite{ubbelohde1939isotope}, i.e., the elongation of the H-bonds upon deuteration, a rather small effect of the order of some hundredths of an angstrom, is well reproduced.
The Ubbelohde effect indicates that quantum fluctuations strengthen the H-bonds when DKDP is replaced by KDP. The fine details of the plots in Fig.~\ref{fig:bonds}(c) are interesting. The small positive slope of $d_{\mathrm{O}...\mathrm{H}}$ with $T$, observed in the simulations of DKDP above and below $T_c$ and only above $T_c$ in KDP, is a manifestation of thermal expansion and indicates that thermal fluctuations weaken the H-bonds, albeit to a rather small extent. The effect can be clearly seen also in the experiments on KDP for $T>T_c$. Interestingly, the effect is absent in the KDP simulations below $T_c$, suggesting that in the ferroelectric phase of this system, quantum fluctuations dominate over thermal fluctuations.

\begin{figure}[h!]
    \centering
    \includegraphics[width=0.65\linewidth]{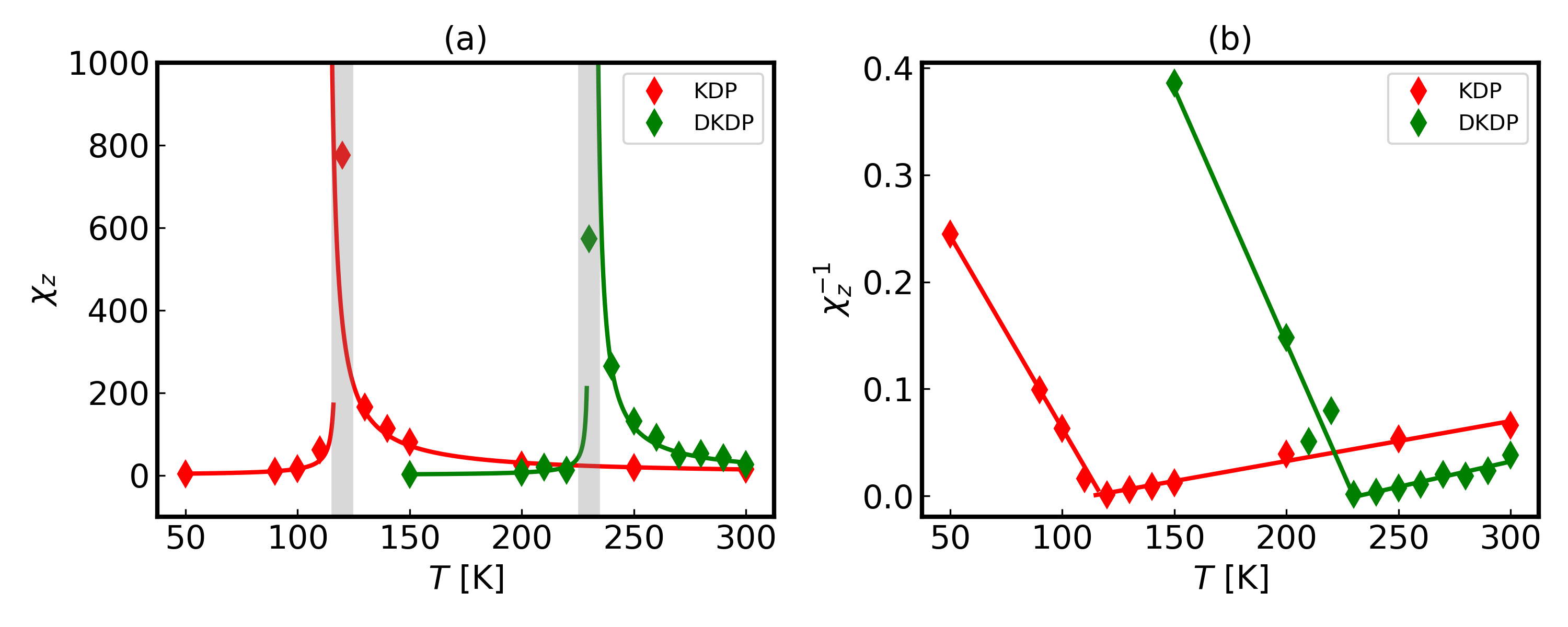}
    \caption{ (a) Predicted susceptibility as a function of $T$. Solid red and green lines are calculated from the inverse susceptibility fitted to the Curie-Weiss law. The vertical thick grey line gives a rough estimate of $T_c$ with uncertainty attributed to finite-size effects. (b) Predicted inverse susceptibility as a function of $T$. Solid lines indicate optimal Curie-Weiss fitting. }
    \label{fig:chi}
\end{figure}

In Fig.~\ref{fig:chi}(a), we report  
the temperature dependence of the longitudinal dielectric susceptibility $\chi_z$ of KDP and DKDP, respectively. $\chi_z$ is found to be in good qualitative agreement with experimental measurements~\cite{chabin1977polarization}. In addition, for both KDP and DKDP, $\chi_z$ displays a discontinuity at $T_c$ as expected for a first-order phase transition. 
Fig.~\ref{fig:chi}(b) plots the inverse susceptibility $\chi_z^{-1}$ as a function of T, which exhibits a typical Curie-Weiss behavior $\chi_z^{-1} \propto T-T_c$ near the phase transition. The optimal linear fitting to Curie-Weiss law is plotted in Fig.~\ref{fig:chi} for $T$ below and above $T_c$, respectively.


\section{Dynamic correlation in proton/deuteron hopping in the ferroelectric phase}

To analyze in more detail dynamic correlation effects in the hopping of protons or deuterons in the ferroelectric phase, we consider the shortest closed loop of H-bonds connecting adjacent phosphate groups.
\begin{figure}[tbh]
\centering\includegraphics[width=0.9\linewidth]{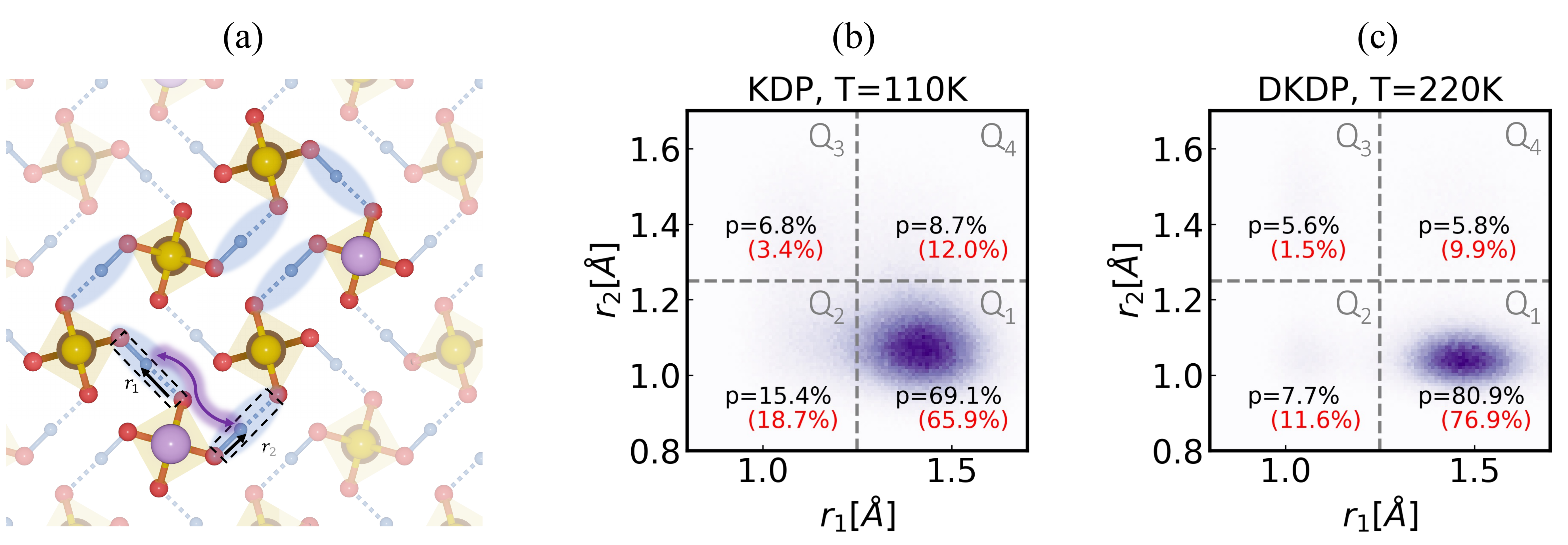}
    \caption{ (a) Schematic representation of the shortest loop of phosphate groups mutually connected by O-H..O bonds. A pair of neighboring hydrogen atoms in the circuit are boxed. The purple wavy line indicates correlation. (b-c) $n(r_1,r_2)$ of KDP and DKDP at different temperatures. The four blocks are divided by $r_1=r^*$ and $r_2=r^*$, as shown by the grey dashed lines. $p$ is the fractional probability of a quadrant, taking correlation in the proton jumps into account, while the red value in parentheses gives the corresponding probability in absence of correlation (see text). }
    \label{fig:nnb_corr}
\end{figure}
In KDP/DKDP this is a 6-fold bond ring, like the one shown in Fig.~\ref{fig:nnb_corr}(a) with the bonds indicated by the light blue ellipses. In the classical ground state, half of the bonds in this ring are oriented clockwise and half anticlockwise, as required for a broken symmetry Ferro environment. A fully ordered ring, clockwise or anticlockwise, would contribute zero to the polarization. Thus, half-ordering of the 6-fold rings can be viewed as the manifestation, in the bond picture, of the spontaneously broken symmetry of the phosphate groups in the ferroelectric phase. Topologically equivalent 6-member ring structures are found in other H-bonded crystals, such as ferroelectric ice XI~\cite{howe1989determination} or antiferroelectric ice VIII~\cite{pruzan2003phase}. To quantify correlation, we consider the two adjacent bonds having length $r_1$ and $r_2$, respectively, in Fig.~\ref{fig:nnb_corr}(a). After averaging over all the pairs of bonds in the simulation box equivalent by translational symmetry to $r_1$ and $r_2$,
we compute the normalized joint distribution $n(r_1,r_2)$ and its marginal distributions $n(r_1)$ and $n(r_2)$ from the configurations of the ring polymers. Plots of $n(r_1,r_2)$ at $T=T_c - 10 \mathrm{K}$ for KDP and DKDP, respectively, are reported in Fig.~\ref{fig:nnb_corr}(b) and (c). Six different pairs of adjacent bonds can be selected in a 6-fold ring like the one of Fig.~\ref{fig:nnb_corr}(a), but their joint distributions are statistically equivalent to $n(r_1,r_2)$. In Fig.~\ref{fig:nnb_corr}(b) and (c) the characteristic length $r^*$, distinguishing donor and acceptor H-bond types, is used to draw the vertical and horizontal dashed grey lines.  These lines divide each figure into four quadrants, corresponding to configurations with $r_1>r^*$ and $r_2<r^*$ (quadrant $Q_1$), $r_1<r^*$ and $r_2<r^*$ (quadrant $Q_2$), $r_1>r^*$ and $r_2>r^*$ (quadrant $Q_4$), $r_1<r^*$ and $r_2>r^*$ (quadrant $Q_3$), respectively. 
The integral of $n(r_1,r_2)$, extended to each quadrant, gives that quadrant's population $p$, expressed as the percentage typed in black in the figure. The integral of $n(r_1)n(r_2)$, extended to each quadrant, gives the quadrant's population in the absence of correlation, reported in the red parenthesis in the figure.
For both $n(r_1,r_2)$ and $n(r_1)n(r_2)$, the fraction of unswitched bond pairs, associated to $Q_1$, is larger in DKDP than in KDP, as expected due to the more prominent role of quantum tunneling in KDP. Meanwhile, the population of $Q_2$ is always larger than, instead of being equal to, $Q_4$, as one would have expected if only asymmetric H-bond structures were present. It is the presence of symmetric H-bond structures that changes the balance, with double donor structures O-H-O much more likely than their double acceptor counterparts O...H...O. 
Correlation accounts for the difference of black and red populations $p$ in the figure. The effect of the correlation is qualitatively similar, albeit slightly more pronounced, in DKDP than in KDP. Single switches of $r_1$ or $r_2$ convert two adjacent Ferro units into two adjacent Takagi units, but correlation reduces the fraction of single switches of $r_1$ ($Q_2$) and $r_2$ ($Q_4$), reducing local charge fluctuations. The reduction of $Q_2$ and $Q_4$ is compensated by the increase of $Q_1$ and $Q_3$, which may be attributed to return switches and double switches, respectively. The population of $Q_3$ is related to the population of Slater defects, because simultaneous switches of $r_1$ and $r_2$  converts a pair of Ferro units to one Slater and one Takagi unit. In DKDP at temperature just below $T_c$, the $Q_3$ population is almost four times larger than its uncorrelated counterpart, suggesting that correlated proton jumps contribute significantly to the population of Slater defects. The effect is weaker in KDP where the $Q_3$ population is merely doubled by correlation at temperature just below $T_c$. 

In the above treatment, we have only considered switches of one or two protons belonging to a pair of adjacent H-bonds. We have not considered the correlated multi-switch processes in ordered H-bond loops that were studied in Ref.~\cite{lin2011correlated}, because the latter require ordered loops that do not contribute to the polarization and, thus, occur very rarely in the ferroelectric phase. They should be present, however, in the paraelectric phase.   


\bibliography{si.bib}